\documentclass[useAMS,usenatbib]{mn2e}
\usepackage{multirow}
\usepackage{graphicx}
\usepackage{footnote}
\usepackage{amsmath}
\usepackage{amssymb}
\usepackage{caption}
\graphicspath{{Images/}}
\DeclareMathOperator*{\argmax}{\arg\!\max}

\title{Modelling multimodal photometric redshift regression with noisy observations}
\author{S.~D. K\"ugler, N. Gianniotis\\
Heidelberg Institute for Theoretical Studies, Schloss-Wolfsbrunnenweg 35,
69118 Heidelberg, Germany}

\begin{document}

\date{}

\pagerange{\pageref{firstpage}--\pageref{lastpage}} \pubyear{2015}

\maketitle

\label{firstpage}

\begin{abstract}
In this work, we are trying to extent the existing photometric redshift
regression models from modeling pure photometric data back to the
spectra themselves. To that end, we developed a PCA that is capable
of describing the input uncertainty (including missing values) in a 
dimensionality reduction framework. With this ``spectrum generator'' at hand, we 
are capable of treating the redshift regression problem in a fully Bayesian 
framework, returning a posterior distribution over the redshift. 
This approach allows therefore to approach the multimodal regression problem 
in an adequate fashion. In addition, input uncertainty on the magnitudes can be 
included quite naturally and lastly, the proposed algorithm allows in principle 
to make predictions outside the training values which makes it a fascinating 
opportunity for the detection of high-redshifted quasars.
\end{abstract}


\section{Introduction}
The exploration of the past development of the universe has been mainly driven 
by the detection and investigation of highly-redshifted extragalactic sources, 
such as the quasi-stellar objects (QSO, \citealp{1993ARA&A..31..473A}). The study 
of the distribution of these objects over space and time allows do draw precise 
conclusions about how the universe has initially formed and developed since then
\citep{2010ASSP...14..255A}. Additonally, photometric redshifts have been used 
in the studies of galaxy clusters \citep{2008MNRAS.387..969A} and in constraining 
the galaxy luminosity function \citep{1996AJ....112..929S}.

Since the detection of the first quasars a significant time of research has been spent
in estimating the redshift, caused by the expansion of the universe, to these ultra-luminous
objects. While spectroscopic surveys are extremely precise in doing so, they are extremely 
time-intensive and can not be used to study a large fraction of the objects known to date.
Instead photometric surveys are used to infer knowledge about the nature and the redshift of 
the quasars. Originally, this was done in a template-based way \citep{2000A&A...363..476B} 
and only rather recently the number of data-driven approaches has increased drastically
(\citealp{2010MNRAS.406.1583W, 2011MNRAS.413.1395O, 2011MNRAS.418.2165L} and many more). In 
these works the main focus has been on the comparison of methodology instead of the introduction 
of new concepts and the community seemed to have agreed on, that the random forest is tailored
for this task in terms of reproducability, precision and computational complexity.

\begin{figure}
\centering
\includegraphics[width=.5\textwidth]{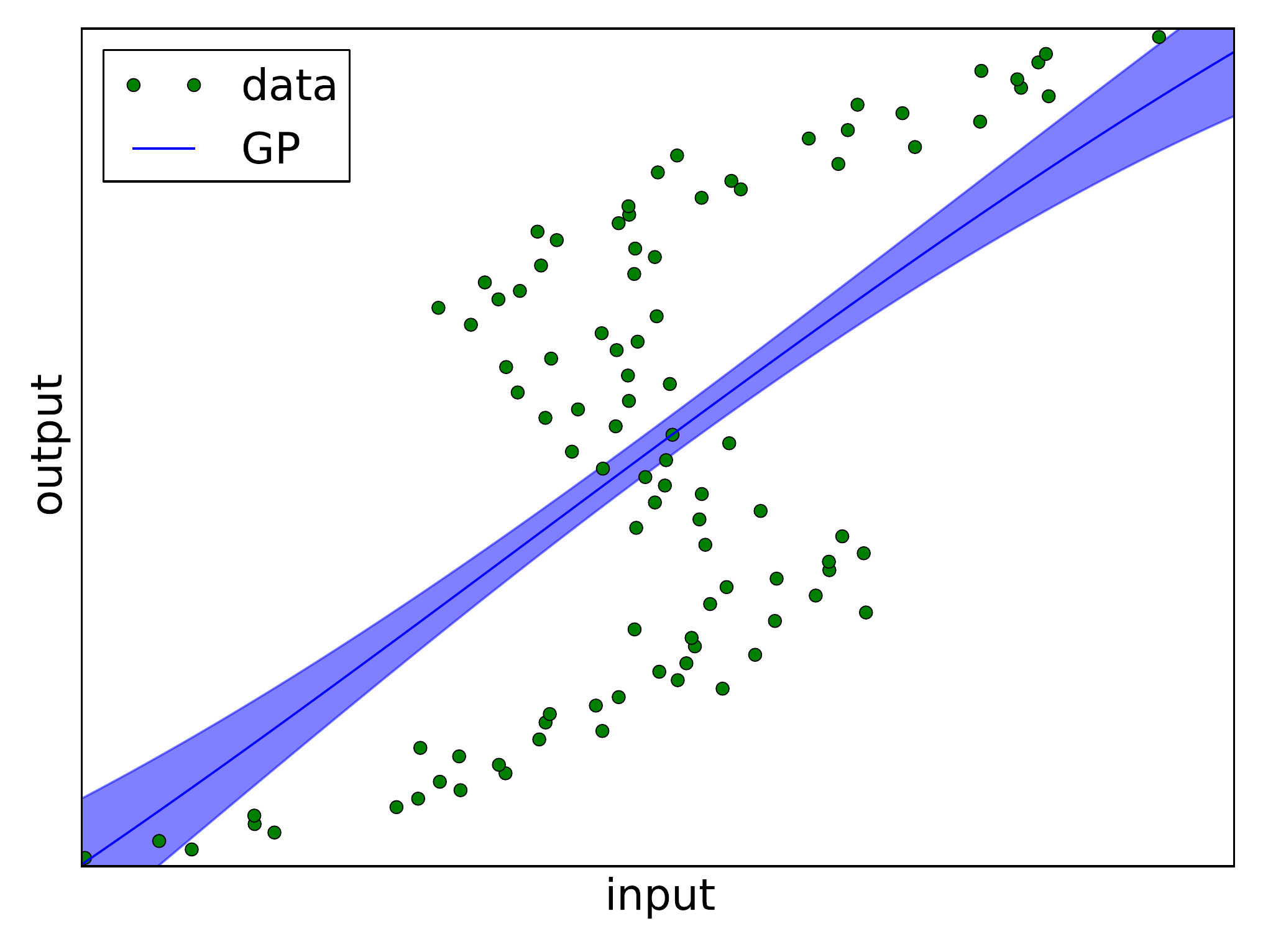}
\caption{Example of a multimodal problem. It is evident that the Gaussian process is not capable of 
describing the multimodal data accordingly.\label{fig:multimodal}}
\end{figure}

In our work, we want to present an algorithm that considers a number of problems in redshift 
regression that have been known to the community for a long time but have not been tackled 
and/or been ignored over the last decade. As it can be study in all plots showing the regressed
redshift versus the actual redshift: the redshift regression problem is actually \emph{multimodal}.
This means that a given color can generally be explained by \emph{more than one} redshift, cf. 
Fig.\,\ref{fig:multimodal}.
In our work, we will also show that the RMS is an inadequate measure to estimate the accuracy of
photometric redshift regression algorithms and present a more useful measure. This measure will 
be based on the posterior distribution over the redshift which should be rather considered in a 
multimodal problem. Another striking problem of the existing methodology is, that the uncertainty
of the input data can not be considered so far, i.e. that uncertainties, or even more drastic missing
values, of the colors can not be considered. Despite that, it has been claimed, that these regression
algorithms can be used for predicting out-of-sample (i.e., higher redshifts than provided by the 
training) regression values. While in practice this might be possible, it is conceptually highly 
questionable whether this is the right concept. Instead we can provide with the model-based approach
an alternative in the search for highly-redshifted extra-galactic objects, however, a lot more work 
has to be done in order to achieve this goal.

We start with a description of the data used in \mbox{Section\,\ref{sec:data}}, followed by a 
description of the methodology
in \mbox{Section\,\ref{sec:methodology}}. In the following part the results are presented 
which contains also a direct comparison with the random forest. Subsequently, we summarize 
in \mbox{Section\,\ref{sec:discussion}} the work and show potential prospects for continuing 
the presented work.

\section{Data}
\label{sec:data}
\begin{figure}
\centering
\includegraphics[width=.5\textwidth]{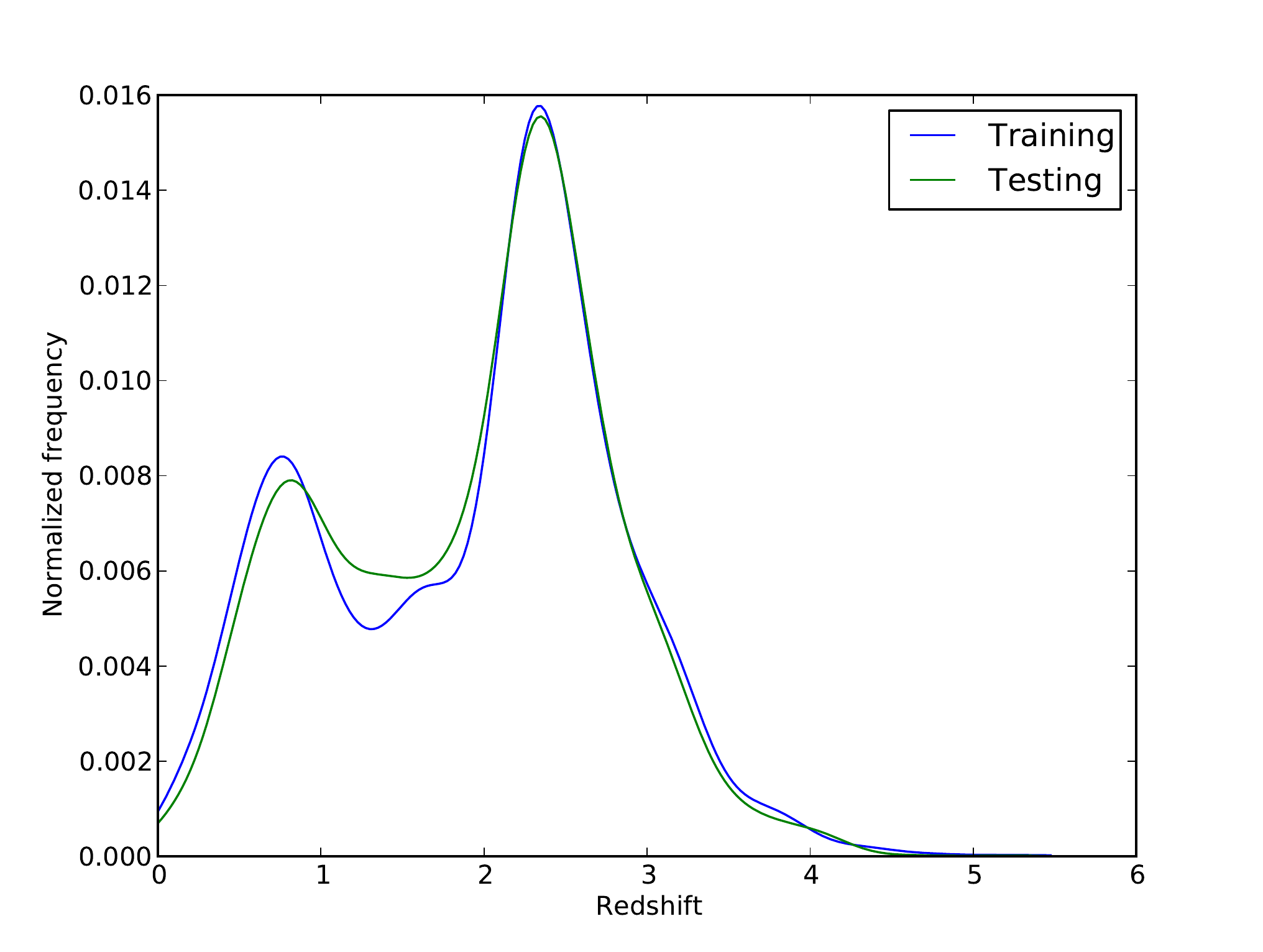}
\caption{Redshift distribution of training and test data.\label{fig:redDist}}
\end{figure}

The presented methodology is conceptually new and is very different from the approaches
used in the literature of photometric redshift regression. For this reason, we demonstrate
our methodology on a small subset of quasars contained in the BOSS catalog. First, we 
extract 7506 randomly selected quasar spectra from BOSS which we divide into a training (5000)
and test set (2506), the redshift distributions can be seen in Figure\ref{fig:redDist}. The 
idea is now to extract the photometry directly from the spectra instead of using their 
observed direct photometric counterparts. This way of approaching the problem has many 
advantages:
\begin{itemize}
\item no calibration of the zero points needed
\item no uncertainties in the observables (spectra are considered noiseless)
\item full control over how data have been generated
\end{itemize}
One of the downsides of using the spectra is that not the entire $u$ band is covered
and thus we just have 3 colors to our availability (in the presented case these will be 
the three independent colors $g-r$, $g-i$, $g-z$). Note that in our methodology the fluxes
themselves are used instead of colours and therefore the data presented to our algorithm are
four dimensional. This is not an advantage as our model contains an additional scaling that
has to be optimized. 
\subsection*{Preprocessing}
All the required spectra are downloaded from the SDSS 
server.
In a first step,
all the spectra are binned with a binning factor of 10 according to the following rules:
$$\lambda^j_{new} = \frac{1}{10}\sum_{i=10j}^{10(j+1)}\lambda^i_{old}$$
$$f^j_{new} = \frac{\sum_{i=10j}^{10(j+1)}f^i_{old} (\Delta f^i_{old})^{-2}}{\sum_{i=10j}^{10(j+1)}(\Delta f^i_{old})^{-2}}$$
$$\Delta f^j_{new} = \frac{1}{\sum_{i=10j}^{10(j+1)}(\Delta f^i_{old})^{-2}}$$
where $\lambda$, $f$ and $\Delta f$ are the wavelength, the spectral flux and the 
error of the spectral flux respectively. Subsequently, the spectra are shifted into 
their restframe and the flux values are extracted on a fixed grid ($\lambda \in [500,10400]$ in 1000
equally spaced steps) using spline interpolation. All missing parts are highlighted by an infinite 
flux and an infinite error on the flux.

\section{Methodology}
\label{sec:methodology}
\subsection{Uncertain PCA}
\subsection{Redshift regression model}
In the previous section, a detailed description of the principle of the PCA, fed
with uncertain spectra, was given. This PCA provides us now with a tool that can
generate any kind of observed spectrum, simply based on the low-dimensional
coordinates. In order to convert this into a redshift regression model, we
assume the following things: With the PCA-weights $W\in \mathcal{R}^{LxD}$ and
given coordinates $\theta \in \mathcal{R}^{1xL}$ we obtain a spectrum
$\mathcal{S} = \theta \cdot W \in \mathcal{R}$, with $\cdot$ denoting the dot
product. In general, this spectrum is not necessarily at a redshift of $z=0$,
and thus the spectrum would be a function of $\mathcal{S}(\theta, z)$. Finally
we are observing photometric magnitudes and therefore, we have to integrate the
obtained spectrum over the filter curve of a filter $b$. The flux
$\mathcal{I}$ in a band b is thus computed as
$$\mathcal{I}_b(\theta,z) = \frac{\int_0^\infty \lambda S(\lambda/(z+1))
f_b(\lambda)d\lambda}{\int_0^\infty \lambda 
f_b(\lambda)d\lambda}.$$
Since the spectra are discrete the transformation \mbox{$S(\lambda/(z+1))$} is not
helpful for calculating the gradient with respect to $z$. That is why, we use
the replacement
$$\tilde{\lambda} = \lambda/(z+1)$$
$$\frac{d\tilde{\lambda}}{d\lambda} = 1/(z+1) \Rightarrow
d{\lambda}= (z+1)d\tilde{\lambda}$$
and thus
\begin{equation}
\begin{split}
\mathcal{I}_b(S,z) & = \frac{\int_0^\infty (z+1)\tilde{\lambda}
S(\tilde{\lambda})
f_b((z+1)\tilde{\lambda}) (z+1)d\tilde{\lambda}}{\int_0^\infty
(z+1)\tilde{\lambda} f_b((z+1)\tilde{\lambda})(z+1)d\tilde{\lambda}} \\
& =
\frac{\int_0^\infty \lambda
S(\lambda)f_b((z+1)\lambda)d\lambda}{\int_0^\infty
\lambda f_b((z+1)\lambda)d\lambda}.\\
\end{split}
\end{equation}
Effectively, we have now pushed the redshift from the discontinuous spectrum to
the discontinuous filter bands by replacing $\tilde{\lambda} \rightarrow
\lambda$. However, these can be easily approximated with an analytical function,
here with a mixture of Gaussians $$f(\lambda)=\sum_{c=1}^{N_{comp}}V_c\exp
\left(-0.5\left(\frac{\mu_c-\lambda}{\sigma_c}\right)^2\right)$$ with
$V_c$,$\mu_c$,$\sigma_c$ being the weights, the means and the widths of each of
the $N_{comp}$ Gaussian components. 

In order to compute now the expected flux from our model, we can
approximate this integral with a regular Riemann sum, where the bin width
$\Delta$ is given by the distance between two regularly sampled grid points, as
described in the preprocessing. Finally, the flux in band $b$ is
computed as 
\begin{equation}
\begin{split}
\mathcal{I}_b(\theta,z) & \approx \frac{\Delta\sum^D_d \lambda_d
\theta\cdot W(\lambda_d) f_b((z+1)\lambda_d)}{\Delta\sum^D_d
\lambda_d f_b((z+1)\lambda_d}) \\
& = \frac{\sum^D_d
\lambda_d \theta\cdot W(\lambda_d) f_b((z+1)\lambda_d)}{\sum^D_d
\lambda_d f_b((z+1)\lambda_d)}.\\
\end{split}
\end{equation}
In summary, we know how the flux in a band $b$ for a spectrum defined by PCA
coordinates $\theta$, redshift $z$ can be computed. Now, all we have to do is to
convert the observed magnitudes to equivalent fluxes in the
spectra\footnote{Note that we prefer to work in flux space, as the PCA might
well return also negative spectra, which are unphysical, but can still occur as
part of the optimization process. Taking the log of such a negative spectrum
would return infinity in the loss function.}. 


Lastly, an arbitrary scaling constant $s$ has to be introduced, in
order to accomodate for the difference in average flux and thus the full loss
function $\mathcal{L}$ reads
\begin{equation}
\mathcal{L}\left(\theta, z, s\right) = \sum_b
\left( \frac{s\mathcal{I}_b\left(\theta,z\right) -
10^{-0.4(T_b-ZP_b)}}{\sigma_b} \right) ^2 
\end{equation}
where $T_b$ and $\sigma_b$ are the magnitudes and their respective uncertainties
of the object observed in band $b$. For this objective loss function, the
gradient with respect to $s$, $\theta$ and $z$ are computed and the
loss is optimized using conjugate gradient (CG). 

In experiments we learned that 
even a PCA with just 10 principle components leads to severe over-fitting of the
likelihood function. For this reason, the optimizer converges to the closest local
minimum (which due to the over-fitting is one of many local minima) during the 
non-convex optimization of the redshift. It is quite estonishing to see that a 
10-dimensional linear model can lead to such behaviour but it seems that the 
broad integration over the filters washes out too many details which can be imitated
by the PCA. The models that are fitted are usually unphysical since the models 
allowed by the PCA are not necessarily supported by data. In order to counteract 
this problem, we could either constrain the coordinates in the lower-dimensional 
space to the regions supported by actual data or draw prototypes from the projected 
data. As the former method would require even more assumptions about flexibility of the
model, we decided to draw prototypes instead. Apart from imputing less assumptions, 
that puts us into the situation to evaluate redshifts in a fully \emph{Bayesian} 
framework.

To that end all the parameters from the model are evaluated on a grid, see e.g. 
\mbox{Tab.\,\ref{tab:params}}. For the scaling\footnote{Note that the scaling could be omitted by 
optimizing colors instead of bands, then of course the input dimension would decrease 
by one accordingly.} we assume an uninformative prior
$$Pr(s) = \frac{1}{\log s_{max}-\log s_{min}}\frac{1}{s}$$
while for the redshift and the prototypes a uniform prior is assumed respectively. 
Finally, we can compute the posterior of the model, given data, over $z$ as  
\begin{equation}
P(z) = \frac{\sum_{s} \sum_{p} 
\exp{\left(-0.5\mathcal{L}\left(p, z, s\right)\right)} Pr(z)Pr(s)}
{\sum_{z}\sum_{s} \sum_{p} 
\exp\left(-0.5\mathcal{L}\left(p, z, s\right)\right) Pr(z)Pr(s)}.
\label{eq:posterior}
\end{equation}
With the computation of this posterior we can achieve multiple things. First of all,
the model is aware of the multi-modality of the problem, i.e., with the given data 
we obtain an a posteriori density of an \emph{arbitrary} shape. In the results section,
we will show how this can be effectively used to estimate redshifts and compare this 
to classical regression algorithms like the random forest. Secondly, we have now the 
choice to select a prior on $z$. In the given task, we chose an uninformed prior 
(uniform). This is in strong contrast to the machine learning architecture usually 
applied in this case. There, the bias which is inherent in the \emph{training} is 
automatically propagated to the predictions as well. While this might be of advantage 
in some cases, it is a generally unwanted side-effect of the training procedure. In 
section XX we will explicitly make use of this advantage and show that the detection
of high-redshifted quasars is more favorable in an unbiased setting.

\begin{table}
\centering
\begin{tabular}{lll}
parameter & grid & grid points \tabularnewline
\hline
scaling & 0.5:2.0:0.025 & 60 \tabularnewline
prototypes & 1:5000:1 & 5000 \tabularnewline
redshifts & 0:5.5:0.025 & 220 \tabularnewline
\end{tabular}
\caption{Evaluation parameters for the Bayesian framework\label{tab:params}}
\end{table}

\subsection{Application of the algorithms}

We apply our method on the 2506 training objects, where the templates are the 5000 
PCA reconstructed training spectra. The dimensionality of the PCA embedding is 
(arbitrarily) set to 10. In order to calculate the posterior distribution according to
Eq.\,\ref{eq:posterior} we need to choose the variances ${\sigma_b}^2$ for all the bands $b$.
In principle this error is dominated by the uncertainty of the photometric measurements, 
however in the presented work, we artifically created them by integrating the raw spectra 
and thus no uncertainty enters the photometry. Still we see a deviation between the real 
spectra and the recovered ones which is due to the imperfect reconstruction of the spectra
using the PCA. So in this case we use the training set to estimate the standard deviation 
for each band by comparing the magnitude obtained from the real and the reconstructed 
spectrum.  

The mode prediction $z_{mode}$ is computed by finding the 
maximum of the posterior distribution
$$z_{mode} = \argmax\limits_{z}(P(z)).$$
These mode values are then considered the predictions of our regression model from 
which the $MAD$ and the $STD$ are computed.

The random forest \citep{Breiman:2001:RF:570181.570182} is trained on the 5000 training 
objects and used for prediction on the 
test values. The number of trees is set to 1000, even though the data are just three 
dimensional. In order to convert the point predictions of the random forest into a 
probability density, we assume that the predictions made by the random forest $z_{RF}$ 
are actually normally distributed as $P(z)=\mathcal{N}\left(z|z_{RF},\sigma^2\right)$. 
Consequently, we have to determine a value for $\sigma$. The most intuitive way is to 
compare actual and predicted values for the training dataset, leading to $\sigma=0.30$. 
Alternatively, $\sigma$ can be chosen such that $\frac{1}{N}$logL$_{\mathrm{True}}$ is
optimized ($\sigma=0.76$). However, it should be noted, that this is not a practical way
of optimization, as by definition test data should not be used for the optimization of model
parameters. Here, this is just done for convenience to show that even a very uncertain 
prediction can not outperform the presented algorithm.

The Gaussian process \citep{Rasmussen} is chosen as a regression model, as it provides immediatelly a 
(normal) distribution of the output values. Therefore, these mean predictions and their 
variances can be directly used to calculate $\frac{1}{N}$logL$_{\mathrm{True}}$ and
$\langle KLD \rangle$.

\section{Results}
\label{sec:results}
\subsection{Validation}
A common measure to compare the performance of the redshift regression algorithms is to
use the root mean square (RMS) on the normalized redshift deviation
$$\Delta z_{norm} = \frac{z_{reg}-z_{true}}{1+z_{true}}.$$
Here, we use for convience the standard deviation (STD) which is for $N=2506$
this is also equivalent to the RMS. Another frequently used measure is the median absolute 
deviation (MAD) as this is less susceptible to objects deviating extremely from the locus.

With the example in the introduction we could, however, convincingly show that these measures 
are rather meaningless in the setting of a multimodal model as the optimization for the mean 
behaviour produces misleading predictions. For this reason, we propose two new measure that 
are capable to take into account the multimodality of the regression value. Since our model 
returns a posterior distribution over $z$, we might ask how probable the true redshift $z_{true}$
is according to our model and thus we can compute
\begin{equation}
\frac{1}{N}\mathrm{logL}_{\mathrm{True}} = \frac{1}{N_{test}} \sum_{n=1}^{N_{test}}P_n\left(z_{n;\,true}\right)
\label{eq:logL}
\end{equation}
where $N_{test}=2506$ is the number test items.

Another measure which tells us whether we captured the multi-modality of the data correctly
is to investigate what other redshifts originate from similar colors. For this reason, we 
perform a $k$ nearest neighbour search in the colour space and identify the closest $k$ 
redshifts from the training data. In this case, we chose $k=10$ but the statement also holds
for other values of $k$. On these 10 redshifts we perform a kernel density estimation 
(KDE)\footnote{also a different choice of the bandwidth does not alter the results for the 
discussion, here we 
set the bandwidth to 0.1} and can conequently compute the Kullback-Leibler-divergence between the 
neighboring redshifts $K(z)$, computed from the KDE and the posterior $P(z)$ as 
\begin{equation*}
\begin{split}
KLD(K,P) & = \int K(z) \log{\frac{K(z)}{P(z)}}dz \\
& \approx \frac{1}{k}\sum_{k^{\prime}=1}^k \log{K\left(z_{k^{\prime}}\right)} 
- \log{P\left(z_{k^{\prime}}\right)} \\
\end{split}
\end{equation*}
This is done for all $N_{test}$ objects and the average 
\begin{equation}
\langle KLD \rangle = \frac{1}{N_{test}}\sum_{n=1}^{N_{test}}KLD_n
\label{eq:KLD}
\end{equation}
over them is used. The KLD tells us how dissimilar the two distributions are and 
thus lower values indicate a better agreement.

\begin{figure}
\centering
\includegraphics[width=.5\textwidth]{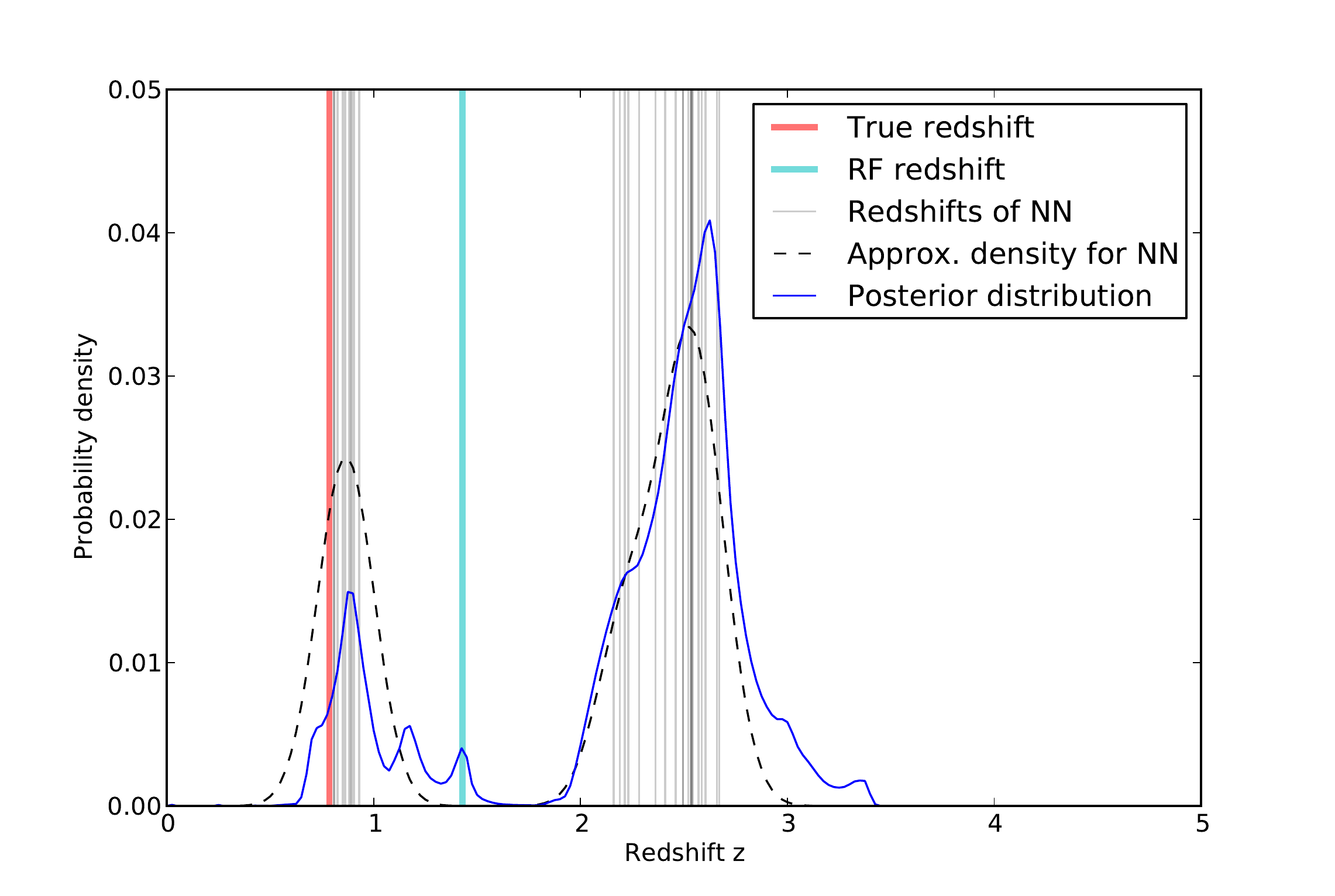}
\caption{For a given data item the actual (spectroscopic) redshift is shown in red. The redshift obtained by random forest regression (or kNN regression) is shown in magenta. In addition, the redshifts of the nearest neighbor objects in terms of the colour space are shown in transparent black, vertical lines. This density is then approximated by a Kernel density estimation (KDE) and compared to the posterior distribution of the presented model.\label{fig:example}}
\end{figure}

At this point the question might arise, why we should not use the $k$ nearest neighbours 
($k$NN) approach to model the posterior density. Inarguably, this would provide us with 
a sampled version of the true posterior distribution, but to actually model the point 
estimates as a distribution, a density estimator would have to be quantified. These 
can estimate the underlying density more accurately the higher $k$ is chosen, however, it
has to be validated what a meaningful value for $k$ is. Therefore, this method would again
depend on several free parameters which may alter the appearence of the distribution 
accordingly. In addition, this would again be a data-driven approach which is neither 
capable of including the uncertainties of the colors nor allows for predictions outside 
the training sample. For this reason, it is only used for cross-validating our developed 
model and to verify that our obtained distribution is similar to the true underlying one. 
In Fig.\,\ref{fig:example} an illustrative example is shown. The red line denotes the actual
redshift of the object, the cyan one the prediction by the random forest. The gray lines are 
the redshifts of 20 nearest neighbours in color space. This distribution is modeled using kernel
density estimation and returns the dashed black curve. Finally, the blue curve is the posterior 
distribution as obtained by our model\footnote{Even though the distribution looks continuous it is a 
discrete distribution which is converted into a piece-wise constant distribution.}
. Clearly, a very similar behaviour is observed, 
strenghtening the principle idea of our model.

\subsection{Comparison of the algorithms}
\begin{figure}
\centering
\includegraphics[width=.5\textwidth]{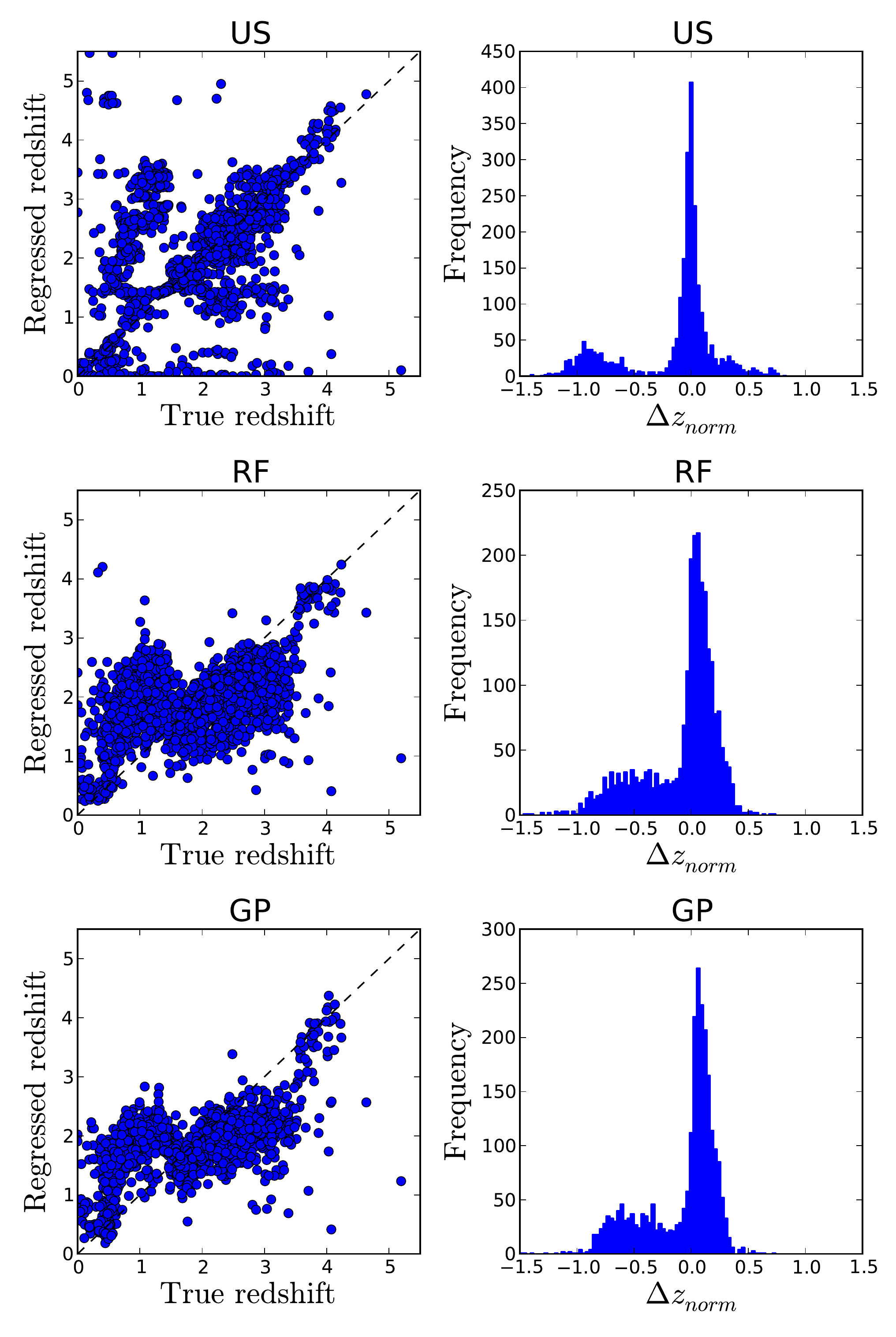}
\caption{Each row shows the results of the experiment noted on top of each plot (our algorithm, random forest and Gaussian process). On the left side the value obtained by the regression algorithm is plotted against the actual value. On the right one can see a histogram of $\Delta z_{norm} = \frac{z_{reg}-z_{true}}{1+z_{true}}$. One can clearly see that for our presented method the peak is much sharper, but the left wing is much more pronounced than for the other two algorithms.\label{fig:allResults}}
\end{figure}

In Tab.\,\ref{tab:results}, the result for the described experiment is shown. For the commonly
used $STD$, we can see that the GP and RF perform comparably well and significantly better than 
our presented algorithm. However, for all remaining measures the presented algorithm performs 
drastically better than the compared algorithms.  
\begin{table}
\centering
\begin{tabular}{lrrrr}
& Here & RF  & RF & GP \tabularnewline
&      & ($\sigma=0.30$) & ($\sigma=0.76$) & \tabularnewline
\hline
$STD$ & 0.476 & 0.344 & 0.344 & \textbf{0.326}\tabularnewline 
$MAD$ & \textbf{0.078} & 0.123 & 0.123 & 0.111 \tabularnewline
$\frac{1}{N}$logL$_{\mathrm{True}}$ & \textbf{-0.514} & -2.979 & -1.138 & -77.875\tabularnewline
$\langle KLD \rangle$ & \textbf{1.082} & 3.491 & 1.536 & 81.660 \tabularnewline
\end{tabular}
\caption{Summary of the results for the presented toy example.\label{tab:results}}
\end{table}
In order to understand the results, we have to understand the difference between the 
measures presented. In Fig.\,\ref{fig:allResults}, the regressed redshift is plotted against
the true one and additionally a histogram over $\Delta z_{norm}$ is shown for each of the algorithms.
On a first look, we can clearly see, why the $STD$ is much worse for the algorithm presented in this
work. While for the GP and RF the points are all distributed along the line, there are some very drastic
deviations apparent in our algorithm. This behaviour is actually a \emph{bias} in the training dataset 
which for the GP and RF is propagated through the algorithm, while in our case it is omitted. Since most
training redshift are located between $[2,3]$ (cf., Fig\,\ref{fig:redDist}), predictions of objects that 
are not very certain will by design between $[2,3]$ as well. This is reasonable, if the redshift 
distribution is \emph{similar} for training in testing. If this can not be guaranteed (as in most 
realistic settings, since the observational biases between surveys can be different), this will effectively 
lead to an amplification of this bias and thus even worse predictions will be produced, as shown later on.
In our model, all redshifts are equally likely (uniform prior), but of course we can influence this 
behaviour by telling our prior about the distributions of redshifts. If we do so, the predictions are, 
by design, much more constraint towards a straight line, which is also reflected in significantly lower
$STD$ of 0.400.

\begin{figure}
\centering
\includegraphics[width=.5\textwidth]{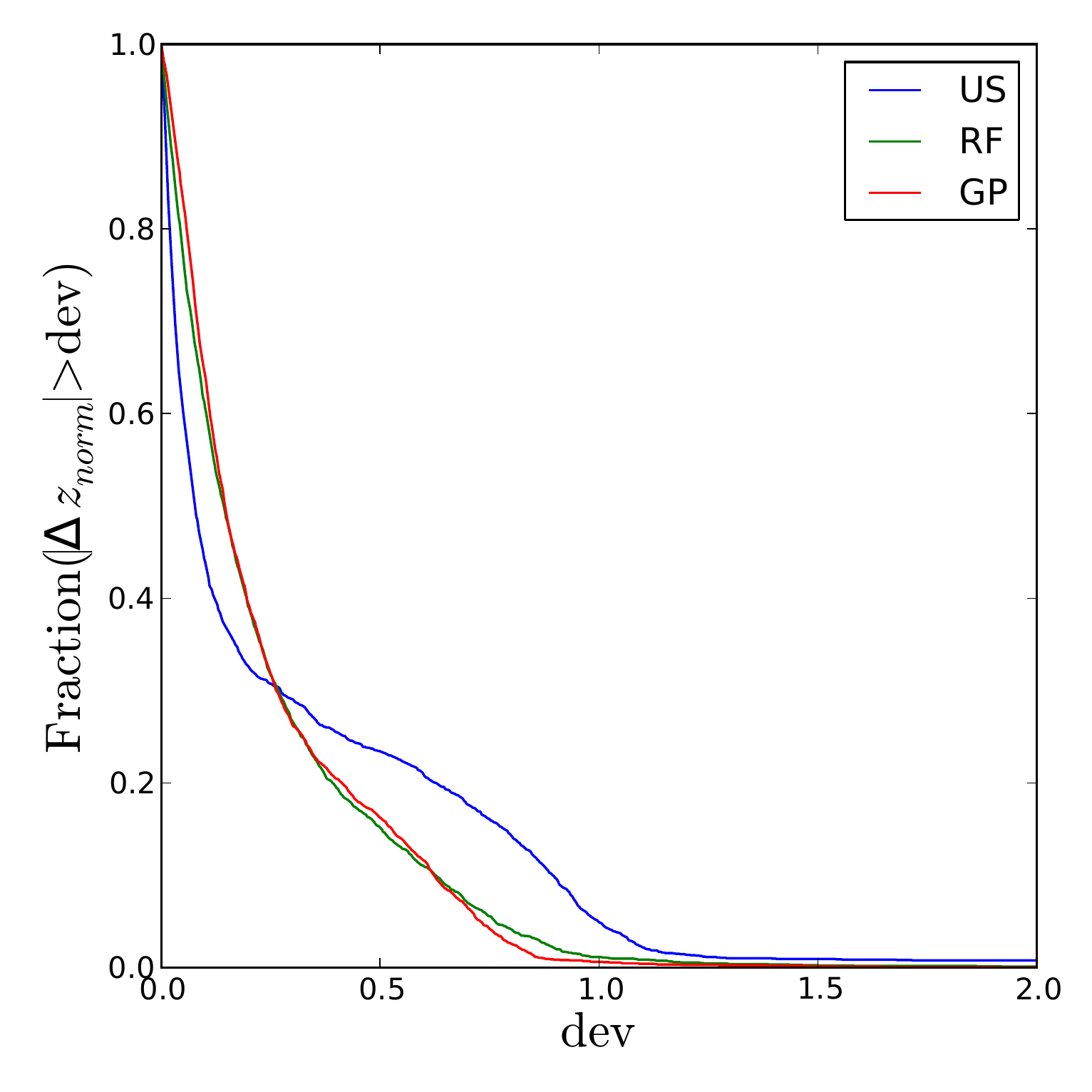}
\caption{The fraction of objects deviating with more than $|\Delta z_{norm}|>dev$ is plotted. While the random forest and the Gaussian process show a very similar behaviour, our presented algorithm can predict the redshift for 60\% of the objects significantly more precise but is heavily influenced from redshifts deviating more than 0.3.\label{fig:fractions}}
\end{figure}

The $MAD$ behaves differently as for this heavily deviating objects are only marginally considered. As a 
consequence, the $MAD$ measures rather the width of the central distribution than the width of the full
distribution. As evident from Fig.\,\ref{fig:allResults}, the predictions obtained from our model are much
more precise than the ones by the RF or GP. This becomes even clearer if we consider the fraction of objects
that deviate more than a certain value, cf. Fig.\,\ref{fig:fractions}. For the vast majority of the objects 
($\approx 70\%$) the deviation from the true value is considerably lower than for the RF and GP predictions. 
The higher precision originates from the superior knowledge provided to the model-based algorithm which 
can describe the behaviour of the spectra in more detail than the data-driven approaches ever could.

As stated in the former sections, it is questionable why for a multimodal problem the average prediction 
should be considered, as this leads necassarily to wrong prediction. Instead, it should be considered how
well the obtained distributions reflect the underlying multimodality of the model. For this reason we measure
the averaged likelihood that each of the models can explain the true redshift of a given item. If we just 
consider the RF optimized on training data ($\sigma=0.3$), the Bayes factor ($\approx 11.8$) provides strong 
evidence that our model is better supported by the provided data. In addition, we can also see that the average 
$KLD$ is by more than a factor of three smaller than the one by the random forest, supporting the power of the 
presented algorithm. In both cases, the GP does not even come to comparable values as the individual predictions
for the GP are appearently over-certain as compared to their true nature.  

\section{Discussion}
\label{sec:discussion}
In this work, we presented an alternative way of modeling photometric redshifts. The idea 
was to combine the advantages of model-based approaches (full model control, Bayesian framework,
multimodality) with the ones from the data-driven approaches (no explicit model-formulation). 
For this reason a probabilistic PCA was developed which can produce projections in the presence 
of missing and uncertain data. In a first step, we tried to utilize this projection to optimize for
the coordinates of the PCA, but learned that if the coordinates are unconstrained already a 10-dimensional
PCA can produce unphysical spectra which then lead to overfitting. Instead, we used the provided reconstructed
spectra themselves as prototypes and therefore converted the continuous model into a discrete one. 
With these prototypes at hand, we were able to treat the regression problem in a fully Bayesian framework
which allowed us to compute the posterior distribution over the redshift.

We compared the distribution of our model with predictions made by the random forest and the Gaussian process.
Even though, we emphasize that this comparison is fairly meaningless as the different approaches are tailored
for different tasks. The standard regression models (RF, GP) are capable of describing a function-like behaviour
(one input maps to a redshift), while it is long known that the photometric redshift regression problem
is a multimodal one (one input can map to several redshifts). Therefore, also the measure of the $RMS$ and the 
$MAD$, as commonly done in the literature, are questionable measures of the quality of a regression algorithm.
For this reason, we introduced two additional measures which reflect the likelihood of capturing the true underlying
redshift and on the other hand a measure denoting how well we can describe similar colors with different redshifts.
By design, our presented algorithm is capable of describing this behaviour, while the standard regression models
are not, as they expect a one-to-one (or many-to-one) mapping. We would like to highlight, that the choice of the 
methodology is only of minor importance, as long as the wrong objective is measured. To treat this multimodality 
purely with data-driven methodology is quite complicated, even though, some approaches already exist (e.g.,
mixture density networks). Consequently, past publications have utilized a measured that simply not 
adequate for the task at hand.
\subsection*{Prospects}
The presented approach can be developed further in two aspects: the methodological one and the astronomical one.
The biggest disadvantage of the current design is, that rather prototypes are extracted and used for prediction
than a continuous PCA model. The problem of the continuous PCA model is its high flexibility which leads to severe
overfitting. For this reason, it would be desirable to constrain the PCA in the lower dimensional space by, for 
example, estimating the density of the spectra. This density estimate could then be used as a prior weighting the
frequency of the a PCA coordinate in the real spectra. In case that this density can be written as a mixture of 
Gaussians, it might be even possible to marginalize the posterior over the PCA coordinates analytically. This
would make the model extremely fast and would omit the point of sampling or drawing prototypes.

From an astronomical point of view, more work has to be done. So far we have just provided the functioning 
on a small toy dataset where the magnitudes were extracted with the provided filter curves and (known and 
noise-free) zero points were added. We chose this setting, as we wanted to have full control on the model 
and not to be distracted by errorneous and noisy calibrations. It is important to notice that a purely 
data-driven approach can deal with this quite naturally, while the presented algorithm depends heavily on 
the correctness of these calibrations. On the other hand, it is of course also possible to include a given 
uncertainty of the zero points into the model and also this can be cross-validated using a training set.
In summary, a much more detailed understanding of how photometric measurements relate to the spectra is 
required.

Another striking advantage of our model is, that we can include uncertainty of the photometric measurements
directly into the model. This includes also \emph{missing} values which are a common struggle in astronomy
due to the different coverage and depth of the surveys. This is contrast to nearly all data-driven algorithms
which can at maximum handle the input uncertainty by sampling (which is hard for a missing value).

A different prospect of our model would be to extent the PCA model towards the infrared. At the moment,
our coverage above $1\mu m$ is very shallow and thus it would be desirable to retrieve near-infrared to 
mid-infrared spectra of low-redshifted quasars (as otherwise the rest-frame would be in the optical again).
This would allow us to include also infrared data as then the coverage of the prototypes would reach into the 
near infrared. It is important to notice that it does not matter whether the infrared spectra are the same 
objects as the optical ones, it only has to be guaranteed that there is considerable overlap with the 
prototypes as they are now.  

\label{lastpage}
\bibliographystyle{mn2e}
\bibliography{bibliography.bib}
\end{document}